\documentclass{llncs}
\usepackage{amsmath}
\usepackage{amsfonts}
\usepackage{graphicx}
\usepackage{adjustbox}
\usepackage{multirow}
\usepackage{caption}
\usepackage[ruled,linesnumbered]{algorithm2e}
\usepackage{algpseudocode}
\usepackage{url}
\usepackage{cite}

\newcommand{\mathbi}[1]{\textbf{\em #1}}

\title{Wikipedia Vandal Early Detection: \\ from User Behavior to User Embedding}

\author{Shuhan Yuan\inst{1} \and Panpan Zheng\inst{2} \and Xintao Wu\inst{2} \and Yang Xiang\inst{1}}
\institute{Tongji University, Shanghai 201805, China,\\
\email{\{4e66,shxiangyang\}@tongji.edu.cn}
\and
University of Arkansas, Fayetteville AR 72701, USA\\
\email{\{pzheng,xintaowu\}@uark.edu}}

\begin{document}

\maketitle

\begin{abstract}
Wikipedia is the largest online encyclopedia that allows anyone to edit articles. In this paper, we propose the use of deep learning to detect vandals based on their edit history. In particular, we develop a multi-source long-short term memory network (M-LSTM) to model user behaviors by using a variety of user edit aspects as inputs, including the history of edit reversion information, edit page titles and categories. With M-LSTM, we can encode each user into a low dimensional real vector, called user embedding. Meanwhile, as a sequential model, M-LSTM updates the user embedding each time after the user commits a new edit. Thus, we can predict whether a user is benign or vandal dynamically based on the up-to-date user embedding. Furthermore, those user embeddings are crucial to discover collaborative vandals.
\end{abstract}

\section{Introduction}

Wikipedia, as one of the world's largest knowledge bases on the web, is heavily relied on thousands of volunteers to make contributions.
This crowdsourcing mechanism based on the freedom-to-edit model (i.e., any user can edit any article) leads to a rapid growth of Wikipedia. However, Wikipedia is plagued by vandlism, namely ``deliberate attempts to damage or compromise integrity'' \footnote{\url{https://en.wikipedia.org/wiki/Wikidata:Vandalism}}. Those vandals who commit acts of vandalism  damage the quality of articles and spread false information, misleading information, or nonsense to Wikipedia users as well as information systems such as search engines and question-answering systems.

Reviewing millions of edits every month incurs an extremely high workload. Wikipedia has deployed a number of tools for automatic vandalism detection, like ClueBot NG \footnote{\url{https://en.wikipedia.org/wiki/User:ClueBot_NG}} and STiki \footnote{\url{https://en.wikipedia.org/wiki/Wikipedia:STiki}}.  These tools use heuristic rules to detect and revert apparently bad edits, thus helping administrators to identify and block vandals. However, those bots are mainly designed to score edits and revert the worst-scoring edits.

Detecting vandals and vandalized pages from crowdsourcing knowledge bases has attracted increasing attention in the research community \cite{Velasco2012Wikipedia,Kumar2015Vews,Heindorf2016Vandalism}.
For example, \cite{Heindorf2016Vandalism} focused on predicting whether an edit is vandalism. They developed a set of 47 features that exploit both content and context information of users and edits. The content features of an edit are defined at levels of character, word, sentence, and statement whereas the context features are used to quantify users, edited items, and their respective revision histories.
\cite{Kumar2015Vews} focused on predicting whether an user is a vandal based on user edits. The developed VEWS system adopted a set of behavior features based on edit-pairs and edit-patterns, such as vandals make faster edits than benign user, benign users spend more time editing a new page than vandals, or benign users more likely edit a meta-page than vandals.  All the above features empirically capture the differences between good edits and vandalism to some extent and there is no doubt that classifiers (e.g., randomforest or SVM) with these features as input can achieve good accuracy of detecting vandalism.

Different from the existing approaches that heavily rely on hand-designed features, we tackle the problem of vandal detection by automatically learning user behavior representations from their edit sequences. Each edit in a user's edit sequence contains many attributes such as  PageID, Title, Time, Categories, PageType, Revert Status, and Content. We transform each edit sequence into multiple aspect sequences and develop a multi-source long-short term memory network (M-LSTM) to detect vandals.  Each LSTM processes one aspect sequence and learns the hidden representation of the corresponding aspect of user edits. The LSTM as a sequence model can represent the user edit sequence with variable-length as fixed-length real vectors, i.e., aspect representations. We then apply the attention model \cite{Bahdanau2015Neural,Yang2016Hierarchical} to derive the user representation, called user embedding, by combining all aspect representations. The user embedding accurately captures all aspects of an user's edit information. Thus we can use user embeddings as classifier input to separate vandals from benign users.
To the best of our knowledge, this is the first work to use the deep neural network to represent users as user embeddings which capture the information of user behavior for vandal detection.

Our approach has several advantages over past efforts. First, neither heuristic rules nor hand-designed features are needed. Second, while each user may have a different number of edits and each user may have different edit behavior, we map each user into the same low-dimensional embedding space. Thus user embeddings can be easily used for a variety of data mining tasks such as classification, clustering, outlier analysis, and visualization. Third, by using various aspect information (e.g., article title and categories), our M-LSTM is able to effectively capture hidden relationships between different users. The derived user embeddings can be used to analyze collaborative vandals who commit acts of vandalism together to impose big damages and/or evade detection. Fourth, our M-LSTM can naturally achieve early vandal detection and has great potential to be deployed for dynamically monitoring user edits and conducting real-time vandal detection.

\section{Related Work}

Deep neural networks have achieved promising results in image \cite{He2016Deep}, text \cite{Mikolov2013Efficient}, and speech recognition \cite{Graves2013Speech}. The key ingredient for the successful of deep neural network is because it learns meaningful representations of inputs \cite{Bengio2013Representation}. For example, in text area, all the words are trained to represent as real-valued vectors called word embeddings which capture the semantic relations among words \cite{Mikolov2013Efficient}. Then, a neural network model can further combine the word embeddings to represent the sentences or documents \cite{Yang2016Hierarchical}. For image recognition, a deep neural network can learn different levels of image representations  on different levels of the neural network \cite{Zeiler2014Visualizing}. In this work, we propose M-LSTM model to train the representation of users and further use them to predict vandals.

Most work for vandalism detection extracts features, e.g., content-based features  \cite{Velasco2012Wikipedia,Adler2007ContentDriven},  context features  to measure user reputation \cite{Adler2011Wikipedia}, spatial-temporal user information \cite{West2010Detecting}, and then uses those features as classifier inputs to predict vandalism. Moreover, \cite{Mckeown2010Got} utilizes search engine to check the correctness of user edits. However, it is difficult to apply these approaches based on hand-design features to detect  subtle and collaborative vandalism.

Wikipedia vandal detection is related to fake review detection. In~\cite{Bin4}, different types of behavior feature were extracted and used to detect the fake reviews in Yelp. \cite{Lim-cikm-2010} have identified several representative behaviors of review spammers. \cite{Xie-kdd-2012} studied the co-anomaly patterns in multiple review based time series.  \cite{Bin2} proposed approaches to detect fake reviews by characterizing burstiness of review.
There has  been extensive research on detecting anomaly from  graph data \cite{akoglu2010oddball,sun2005neighborhood} and detecting Web ranking spams \cite{Han-kdd-2012}. \cite{Ntoulas-www-2006} have studied various aspects of content-based spam on the Web and presented several heuristic methods for detecting content based spam.   Finding time points at which graph changes significantly given a sequence of graphs has also been studied \cite{papadimitriou2010web}. Although some of above approaches can be used for vandal detection, they are not able to automatically extract and fuse multiple aspects of user edit behaviors.

\section{Multi-source LSTM for Vandal Early Detection}

Our key idea is to adopt multiple LSTMs to transform a variable-length edit sequence into multiple fixed-length aspect representations and further use the attention model to combine all aspect representations into the user embedding. As user embeddings capture all aspects of user edits as well as relationships among users, they can be used for detecting vandals and examining behaviors of vandalism.

\subsection{LSTM Revisited}

\begin{figure}
\centering
\includegraphics[width=0.8\textwidth, height=5.5cm, keepaspectratio]{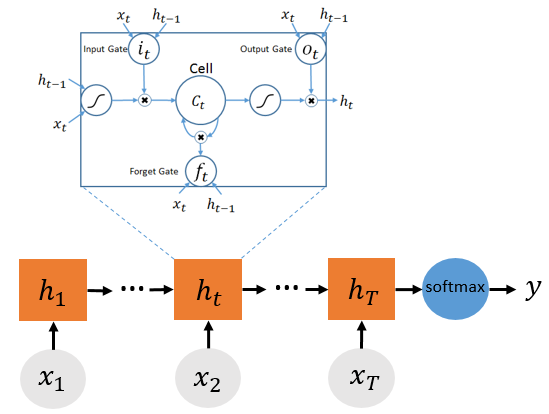}
\caption{Standard LSTM for Classification}
\label{fig:rnn}
\end{figure}

Long short-term memory network \cite{Hochreiter1997Long}, as one class of recurrent neural networks (RNNs), was proposed to model the long-range sequences and has achieved great success in natural language processing and speech recognition recently.
Figure \ref{fig:rnn} shows the structure of the standard LSTM for classification.
Given a sequence $\mathbf{x}=(\mathbf{x}_1, \dots, \mathbf{x}_t, \dots, \mathbf{x}_T)$ where $\mathbf{x}_t \in \mathbb{R}^{d}$ denotes the input at the $t$-th step, LSTM maintains a hidden state vector $\mathbf{h}_t \in \mathbb{R}^{h}$ to keep track the sequence information with the input from the current step $\mathbf{x}_t$ and the previous hidden state $\mathbf{h}_{t-1}$.   LSTM is composed by a special unit called memory block in the recurrent hidden layer to compute the hidden state vector $\mathbf{h}_t$. Each memory block contains self-connected internal memory cells and special multiplicative units called gates to control what kinds of information need to be encoded to the internal memory or discarded. Each memory block has an input gate to control the input information into the memory cell, a forget gate to forget or reset the current memory, and an output gate to control the output of cell into the hidden state. Formally, the hidden state $\mathbf{h}_t$ is computed by
\begin{align}
	\label{eq:lstm}
	\begin{split}
	\mathbf{\tilde{c}}_t &= \tanh (\mathbf{W}_{c} \mathbf{x}_t + \mathbf{U}_{c} \mathbf{h}_{t-1} + \mathbf{b}_c) \\
	\mathbf{i}_t &= \sigma (\mathbf{W}_{i} \mathbf{x}_t + \mathbf{U}_{i} \mathbf{h}_{t-1} + \mathbf{b}_i) \\
	\mathbf{f}_t &= \sigma (\mathbf{W}_{f} \mathbf{x}_t + \mathbf{U}_{f} \mathbf{h}_{t-1} + \mathbf{b}_f) \\
	\mathbf{o}_t &= \sigma (\mathbf{W}_{o} \mathbf{x}_t + \mathbf{U}_{o} \mathbf{h}_{t-1} + \mathbf{b}_o) \\
	\mathbf{c}_t &= \mathbf{i}_t \odot \mathbf{\tilde{c}}_t + \mathbf{f}_t \odot \mathbf{c}_{t-1} \\
	\mathbf{h}_t &= \mathbf{o}_t \odot \tanh(\mathbf{c}_t) \\
	\end{split}
\end{align}
where $\sigma$ is the sigmoid activation function; $\odot$ represents element-wise product; $\mathbf{i}_t$, $\mathbf{f}_t$, $\mathbf{o}_t$, $\mathbf{c}_t$ indicate the input gate, forget gate, output gate, and cell activation vectors and $\mathbf{\tilde{c}}_t$ denotes the intermediate vector of cell state; $\mathbf{W}$ and $\mathbf{U}$ are the weight parameters; $\mathbf{b}$ is the bias term. We denote all LSTM parameters ($\mathbf{W}$, $\mathbf{U}$ and  $\mathbf{b}$ ) as $\mathbf{\Theta}_1$.

After the LSTM reaches the last step $T$, $\mathbf{h}_T$ encodes the information of the whole sequence and is considered as the representation of the sequence. It can then be used as input of the softmax classifier,
\begin{equation}
\label{eq:softmax}
P(\hat y = k | \mathbf{h}_{T}) = \frac{\exp{(\mathbf{w}_k^T \mathbf{h}_{T} + b_k)}}{\sum_{k'=1}^K {\exp(\mathbf{w}_{k'}^T \mathbf{h}_{T} + b_{k'})}},
\end{equation}
where $K$ is number of classes, $\hat y$ is the predicted class of the sequence, $\mathbf{w}_k$ and $b_k$ are the parameters of softmax function for the $k$-th class, and $\mathbf{w}_k^T$ indicates the transpose of $\mathbf{w}_k$. All softmax parameters $\mathbf{W}_k$ and  $\mathbf{b}_k$ over $K$ classes  are denoted as $\mathbf{\Theta}_2$.
The LSTM model aims to optimize  $\mathbf{\Theta}_1$ and  $\mathbf{\Theta}_2$ by minimizing the cross entropy loss function,
\begin{equation}
	\label{eq:loss}
	\mathbi{L} = -\frac{1}{N} \sum_{i=1}^{N} y_i * log(P(\hat{y}_i)),
\end{equation}
where $y_i$ is the true class of the $i$-th sequence, and $N$ is the number of training sequences.

\subsection{Multi-source LSTM}

\begin{figure}
	\centering
	\includegraphics[width=0.8\textwidth]{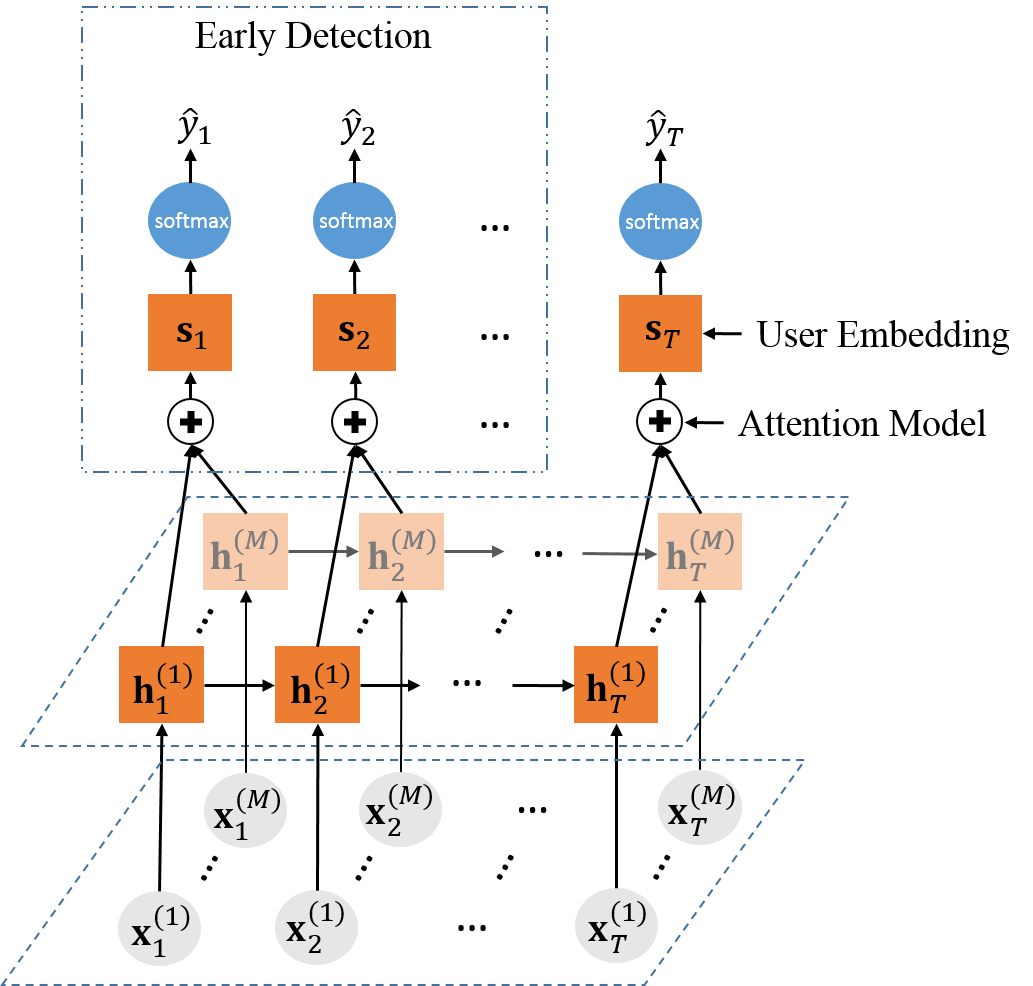}
	\caption{Multi-source LSTM}
	\label{fig:model}
\end{figure}

We develop a multi-source LSTM model to capture all useful aspects of edits. As different aspects carry different weights for vandal detection, we adopt the attention model \cite{Bahdanau2015Neural,Yang2016Hierarchical} to dynamically learn the importance of each aspect. The user embeddings are then used as inputs of softmax classifier to separate vandals from benign users.

Formally, for user $u$ with $T$ edits,  his edits can be modeled as a sequence $\mathbf{e}_u=(\mathbf{e}_{u_1}, \dots, \mathbf{e}_{u_t}, \dots, \mathbf{e}_{u_T})$ where $\mathbf{e}_{u_t}$ includes all related information about the $t$-th edit. Please note different users may have different numbers of edits. For each edit sequence $\mathbf{e}_u$, we transform it into  $M$ aspect sequences.  Its $m$-th aspect sequence, denoted as $\mathbf{x}^{(m)}=(\mathbf{x}_1^{(m)}, \mathbf{x}_2^{(m)}, \dots, \mathbf{x}_{T}^{(m)})$, captures the $m$-th aspect of edit information  and is used as the input of the $m$-th LSTM in our  multi-source LSTM. Figure \ref{fig:model} illustrates our M-LSTM model with $M$ aspect sequences as input.  The last hidden states $\mathbf{h}_T^{(m)}$ ($m=1,\cdots,M$) encode all the aspect information of the user's edit sequence.
We apply the attention model  as shown in Equation \ref{eq:pre_att}--\ref{eq:user} to combine all aspect information into the user embedding.
\begin{align}
	\label{eq:pre_att}
	\mathbf{z}_{T}^{(m)} &= \tanh(\mathbf{W}_{a} \mathbf{h}_{T}^{(m)}),   \\
	\label{eq:att}
	\alpha_{T}^{(m)} &= \frac{\exp(\mathbf{u}_{a}^T \mathbf{z}_{T}^{(m)})}{\sum_{m'=1}^M {\exp(\mathbf{u}_{a}^T \mathbf{z}_{T}^{(m')}})},   \\
	\label{eq:user}
	\mathbf{s}_{T} &= \sum_{m=1}^{M} {\alpha_{T}^{(m)} \cdot \mathbf{h}_{T}^{(m)}},
\end{align}
where $\mathbf{W}_{a} \in \mathbb{R}^{h*h}$ is a trained projection matrix; $\mathbf{u}_{a} \in \mathbb{R}^{h}$ is a trained parameter vector. All the parameters, $\mathbf{W}_{a}$ and $\mathbf{u}_{a}$, in the attention model are denoted as $\mathbf{\Theta}_3$.

In the attention model, we first compute a hidden representation $\mathbf{z}_{T}^{(m)}$ of the last hidden state $\mathbf{h}_{T}^{(m)}$ based on a one-layer neural network by Equation \ref{eq:pre_att}. After obtaining all the $M$ hidden representations, $\mathbf{z}_{T}^{(1)}, \dots, \mathbf{z}_{T}^{(M)}$, we apply the softmax function  to calculate  the weight of each hidden state $\alpha_{T}^{(m)}$ by Equation \ref{eq:att}.  Finally, we compute the user embedding $\mathbf{s}_T$ as the weighted sum of the $M$ hidden states by Equation \ref{eq:user}. The advantage of the attention model is that it can dynamically learn a weight of each aspect according to its relatedness with the user class (e.g., vandal or benign).

We use the user embedding $\mathbf{s}_{T}$ to predict $P(\hat y | \mathbf{s}_{T})$, i.e., the probability of user $u$ belonging to each class $k$ based on softmax function shown in Equation \ref{eq:softmax}.
We adopt the standard cross-entropy loss function (Equation \ref{eq:loss}) to train our M-LSTM model.

Algorithm \ref{algr:method} shows the pseudo-code of M-LSTM training. Given a training dataset $D$ that contains edit sequences and class labels of $N$ users, we first construct the $M$ aspects of edit sequences for each user. After initializing the parameters, $\mathbf{\Theta}_1$, $\mathbf{\Theta}_2$, and $\mathbf{\Theta}_3$, in M-LSTM, in each running, we compute the $M$ last hidden states by LSTM networks (Line \ref{line:hidden}). Then, we adopt the attention model to combine the $M$ hidden states to the user embedding (Line \ref{line:att}). Finally, we update the parameters of M-LSTM by using the user embedding to predict the user label (Line \ref{line:optimize}). The parameters of M-LSTM are optimized by Adadelta \cite{Zeiler2012Adadelta} with the back-propagation.

\begin{algorithm}[h]
	\DontPrintSemicolon
	\SetKwInOut{Inputs}{Inputs}\SetKwInOut{Outputs}{Outputs}
	\Inputs{$D=\{(\mathbf{e}_{u},y_{u});u=1,\cdots,N\}$
			\\ Maximum training epoch $Epoch$}
	\Outputs{Well-trained parameters $\mathbf{\Theta}_1$, $\mathbf{\Theta}_2$, $\mathbf{\Theta}_3$}
	\ForEach {user $u$ in $D$}{
		construct $M$ aspect sequences $\mathbf{x}^{(m)}$ ($m=1,\dots, M$) from the edit sequence $\mathbf{e}_u$;
	}

	initialize parameters $\mathbf{\Theta}_1$, $\mathbf{\Theta}_2$, $\mathbf{\Theta}_3$ in M-LSTM;

	$j \leftarrow 0$;

	\While{$j<Epoch$}{
		\ForEach {user $u$ in $D$}{
			
			compute $\mathbf{h}_{T}^{(m)}$ ($m=1,\dots, M$) on $M$ sequences of aspect vectors;
			\label{line:hidden}
	
			compute the user embedding $\mathbf{s}_{T}$ by attention model (Eq. \ref{eq:pre_att}, \ref{eq:att}, \ref{eq:user}) on $M$ last hidden states;
			\label{line:att}
			
			optimize the parameter $\mathbf{\Theta}_1$, $\mathbf{\Theta}_2$, $\mathbf{\Theta}_3$ in M-LSTM based on the loss function shown in Eq. \ref{eq:loss} with Adadelta.
			\label{line:optimize}
		}

		$j \leftarrow j+1$;
	}
\caption{Multi-source LSTM Training}
\label{algr:method}
\end{algorithm}

{\noindent \bf M-LSTM for Vandal Early Detection}
Our trained M-LSTM model can then be used to predict whether a new user $v$ is vandal or benign given his edit sequence $\mathbf{e}_{v} = (\mathbf{e}_{v_1}, \cdots, \mathbf{e}_{v_t},\cdots)$.
The upper-left region of Figure \ref{fig:model} shows our M-LSTM based vandal early detection.
At each step $t$,  we first derive its $M$ aspect sequences from the user's edit sequence till step $t$. The hidden states are updated with the new input $\mathbf{e}_{v_t}$. Thus, they are able to capture all user's edit aspects until $t$-th step. We then adopt the attention model shown in Equations \ref{eq:pre_att}, \ref{eq:att},  and \ref{eq:user} (replacing all subscript $T$ with $t$) to calculate the user embedding $\mathbf{s}_t$.
The user embedding $\mathbf{s}_t$ captures all the user's edit information till step $t$. Then, we can use the classifier to predict the probability $P(\hat y | \mathbf{s}_t)$ of the user to be a vandal based on $\mathbf{s}_t$. We set a threshold $\tau$ to evaluate whether the user is vandal. When $P(\hat y | \mathbf{s}_t) > \tau$, the user is labeled as vandal.

\section{Experiments}

We conduct our evaluation on UMDWikipedia dataset \cite{Kumar2015Vews}. This dataset contains information of around 770K edits  from Jan 2013 to July 2014 (19 months) of 17105 vandals and 17105 benign users.
We focus on identifying the user behaviors on the Wikipedia articles.
We remove those edits on meta pages (i.e., with titles containing ``User:'', ``Talk:'', ``User talk:'', ``Wikipedia:'') because they do not cause damages.

For each edit, we extract three aspects, article title, article categories, and revert status. We choose these three aspects because both the title and categories capture the topic information of the edited article and  revert status (reported by bots) indicates whether the edit is good or bad. It is imperative to use them to derive user embeddings and then predict whether users are vandal or benign.

We represent article titles and categories to their title embeddings and category embeddings based on word embeddings. Specifically, we first map each word in the titles and categories to its word embedding and then adopt average operation over the word embeddings to get the title embeddings and category embeddings, respectively. The title embeddings and category embeddings reflect the hidden features about the pages. We use the off-the-shelf pre-trained word embeddings \footnote{\url{http://nlp.stanford.edu/projects/glove/}} provided by \cite{Pennington2014Glove}. These word embeddings are widely used and have been shown to achieve good performance on many NLP tasks. We  randomly initialize the words which don't have pre-trained word embeddings.
The dimension of the word embeddings is 50. The dimension of the hidden layer of the M-LSTM network is 32. The training epoch is 25.

\subsection{Vandal Detection}

To evaluate the performance of vandal detection, we split the training and testing dataset chronologically. We use the first 9 months of users as the training dataset and the last 10 months of users as the testing dataset.
The training dataset has 8620 users and the testing dataset has 10418 users.

\begin{table}[]
\centering
\caption{Experimental results on precision, recall, F1, and accuracy of vandal detection with different thresholds}
\label{tb:vandal}
\begin{tabular}{|c|c|c|c|c|}
\hline
$\tau$ & Precision & Recall  & F1      & Accuracy \\ \hline
0.5    & 88.35\%   & 96.67\% & 92.32\% & 91.33\%  \\ \hline
0.6    & 88.69\%   & 96.01\% & 92.20\% & 91.24\%  \\ \hline
0.7    & 89.31\%   & 94.85\% & 92.00\% & 91.10\%  \\ \hline
0.8    & 90.36\%   & 92.27\% & 91.31\% & 90.52\%  \\ \hline
0.9    & 93.13\%   & 74.10\% & 82.53\% & 83.09\%  \\ \hline
\end{tabular}
\end{table}

Table \ref{tb:vandal} shows the precision, recall, F1 and accuracy for vandal detection with different thresholds. Precision indicates the ratio of vandals who are correctly detected. Recall indicates the ratios of vandals who are correctly detected from the test dataset.
The default threshold for binary classification used in vandal detection is 0.5, where our model achieves the best performance. We can also observe that the model achieves good performances of vandal detection with different thresholds $\tau$ from 0.5 to 0.8. Meanwhile, with increasing $\tau$, the precision increases accordingly while the recall decreases, which indicates with a higher threshold, the model can detect vandal more accurate but can mis-classify the vandals as benign users.

We further compare our results with the VEWS approach \cite{Kumar2015Vews}. The VEWS approach uses a set of hand-crafted features to detect vandals. When incorporating  the revision status information, the VEWS can achieve around 90\% classification accuracy. Our M-LSTM achieves better accuracy on vandal detection. More importantly,  our M-LSTM does not need to design dozens of features to predict vandals. Hence our model can be  easily extended to identify vandals from other crowdsourcing knowledge bases like Wikidata.

\subsection{User Embeddings}

As each user has different edits and each edit has many different aspects, it is challenging to derive users' edit patterns. Our M-LSTM derives user embeddings based on user edits. As user embeddings  capture user edit behaviors, they can then be used to differentiate between benign users and vandals and detect collaborative vandals.

\begin{figure}
	\centering
	\includegraphics[width=0.8\textwidth, height=4.5cm, keepaspectratio]{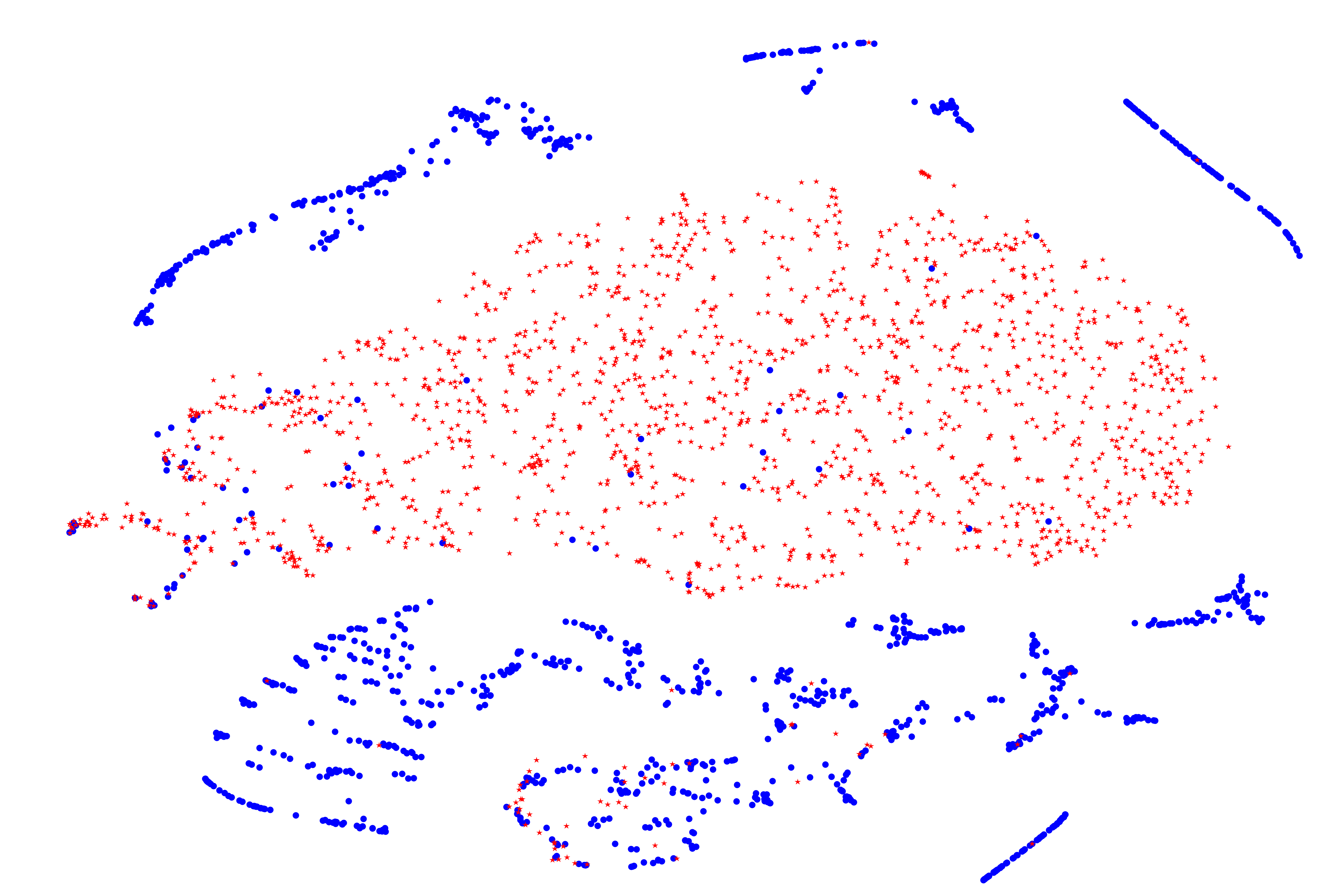}
	\caption{Visualization of 3000 users in the dataset. Color of a node indicates the type of the user. Red: ``Vandals'', blue:``Benign Users''.}
	\label{fig:tsne}
\end{figure}

{\noindent \bf Visualization}
 We randomly select user embeddings of 3000 users and map them to a two-dimensional space based on t-SNE approach \cite{Maaten2008Visualizing}. Figure \ref{fig:tsne} shows the visualization plot of user embeddings from M-LSTM. We observe that the benign users and vandals are clearly separated in the projected space.  This indicates the user embeddings successfully capture the information whether an user is benign or vandal. Hence, they can be used for vandal detection.

{\noindent \bf Clustering of User Embedding}
In this experiment, we adopt the classic DBSCAN algorithm \cite{DBSCANKDD96} to cluster user embeddings. We set the maximum radius of the neighborhood $\epsilon=0.05$ and the minimum number of points $minPts=3$.
DBSCAN produces 211 clusters.  Among them, 139 clusters contain only vandals and the total number of vandals is 502 whereas 46 clusters contain only benign users and the total number of benign users is 495. It indicates the benign users often form large-size clusters. On the contrary, the vandals usually act as small gangs to damage  articles. For the rest 26 clusters that contain mixed vandals and benign users, there are 17 clusters in which the vandals constitute the majority and 9 clusters in which the benign users constitute the majority. Similarly, the  17 (vandal-majority) clusters are small with only 52 vandals whereas there are 3663 benign users in the 9 (benign-majority) clusters. Embeddings of benign users are closer to each other  than that of vandals, which can also be observed in Figure \ref{fig:tsne}.
We conclude that ``Benign users are much more alike; every vandal vandalizes in its own way.''

\begin{table}[ht]
\scriptsize
\centering
\caption{Three example of collaborative vandal groups. The vandals of each group damage the same page(s). Group 1 damages the page ``List of The X Factor finalists (U.S. season 2)'' in 2013-01-05 within a short time window. Group 2 damages the page ``Niels Bohr'' on different days. Group 3 damages the two pages, ``Shakugan no Shana'' and ``The Familiar of Zero'', consecutively in 2014-04-18 within a short time window.}
\label{tb:group}
\begin{tabular}{|c|c|c|c|}
\hline
Group ID                                                                       & User ID                  & Page ID                   & Revision Time \\ \hline
\multirow{8}{*}{\begin{tabular}[c]{@{}c@{}}Group 1\\ 2013-01-05\end{tabular}}  & 4203021                  & \multirow{8}{*}{37310371} & 02:36:32      \\ \cline{2-2} \cline{4-4}
                                                                               & 4203016                  &                           & 02:42:02      \\ \cline{2-2} \cline{4-4}
                                                                               & 4203009                  &                           & 02:42:55      \\ \cline{2-2} \cline{4-4}
                                                                               & 4203006                  &                           & 02:44:58      \\ \cline{2-2} \cline{4-4}
                                                                               & 4202998                  &                           & 02:45:32      \\ \cline{2-2} \cline{4-4}
                                                                               & 4203002                  &                           & 02:47:12      \\ \cline{2-2} \cline{4-4}
                                                                               & 4202988                  &                           & 02:52:12      \\ \cline{2-2} \cline{4-4}
                                                                               & 4202986                  &                           & 02:56:21      \\ \hline
\multirow{3}{*}{Group 2}                                                       & 4584127                  & \multirow{3}{*}{21210}    & 2013-10-04    \\ \cline{2-2} \cline{4-4}
                                                                               & 4597541                  &                           & 2013-10-23    \\ \cline{2-2} \cline{4-4}
                                                                               & 4939865                  &                           & 2014-01-08    \\ \hline
\multirow{10}{*}{\begin{tabular}[c]{@{}c@{}}Group 3\\ 2014-04-18\end{tabular}} & \multirow{2}{*}{5063994} & 2548832                   & 21:33:51      \\ \cline{3-4}
                                                                               &                          & 5982921                   & 21:34:07      \\ \cline{2-4}
                                                                               & \multirow{2}{*}{5063996} & 2548832                   & 21:35:53      \\ \cline{3-4}
                                                                               &                          & 5982921                   & 21:35:53      \\ \cline{2-4}
                                                                               & \multirow{2}{*}{5063998} & 2548832                   & 21:45:06      \\ \cline{3-4}
                                                                               &                          & 5982921                   & 21:45:28      \\ \cline{2-4}
                                                                               & \multirow{2}{*}{5064002} & 2548832                   & 21:47:21      \\ \cline{3-4}
                                                                               &                          & 5982921                   & 21:47:29      \\ \cline{2-4}
                                                                               & \multirow{2}{*}{5064006} & 2548832                   & 21:48:56      \\ \cline{3-4}
                                                                               &                          & 5982921                   & 21:49:01      \\ \hline
\end{tabular}
\end{table}

When setting the maximum radius of the neighborhood $\epsilon = 0$ and the minimum number of points $minPts=2$,  DBSCAN produces 701 groups containing 1687 user embeddings. Note that under this setting all embeddings within a same group are exactly the same.
Among them, 575 groups only contain vandals and the total number of vandals is 1396 whereas  68 groups only contain benign users and the total number of benign users is 144.
The largest vandal group contains 13 vandals and the largest benign group contains 17 benign users.

Table \ref{tb:group} shows three examples of collaborative vandal groups. In Row 1, the group has 8 vandals who attacked the same page consecutively within a short time window.
In Row 2, the group has three vandals who attacked one same page on different days. Because all these vandals were blocked after revising the page, these vandals have high chance to be controlled by a malicious user or group and aim to vandalize the specific page. In Row 3, we show a vandal group containing five vandals. All the five vandals edited the same two pages, ``Shakugan no Shana'' and ``The Familiar of Zero'', which are both Japanese light novels, consecutively within a short time window. These three examples demonstrate one advantage of our M-LSTM, i.e., detecting collaborative vandals with different behavior patterns.

\begin{table}[]
\centering
\caption{Three pairs of vandals and their edited page titles.   Each pair has similar embeddings based on the cosine similarity.}
\label{tb:neighbors}
\begin{adjustbox}{width=0.9\textwidth}
\begin{tabular}{|c|c|c|}
\hline
Vandal IDs                                                     & Page Title                                                                                                                                                                                     & Page Title                                                                                                                                                                                          \\ \hline
\begin{tabular}[c]{@{}c@{}}4266603\\ \&\\ 4498466\end{tabular} & \begin{tabular}[c]{@{}c@{}}Live While We're Young,\\  What Makes You Beautiful, \\ Up All Night (One Direction album), \\ Take Me Home (One Direction album)\end{tabular}                      & \begin{tabular}[c]{@{}c@{}}Live While We're Young, \\ Best Song Ever (song), \\ What Makes You Beautiful, \\ Up All Night (One Direction album), \\ Take Me Home (One Direction album)\end{tabular} \\ \hline
\begin{tabular}[c]{@{}c@{}}4422121\\ \&\\ 4345947\end{tabular} & \begin{tabular}[c]{@{}c@{}}Super Mario 3D World, Super Mario \\ Galaxy, Sonic Lost World, Pringles, \\ Action Girlz Racing, \\ Data Design Interactive\end{tabular}                            & \begin{tabular}[c]{@{}c@{}}Super Mario World, Super Mario World 2: \\ Yoshi's Island, Super Mario Bros. 3, \\ Virtual Boy, Nintendo DS, \\ Kirby Super Star, Yogurt\end{tabular}                    \\ \hline
\begin{tabular}[c]{@{}c@{}}5032888\\ \&\\ 4592537\end{tabular} & \begin{tabular}[c]{@{}c@{}}Matthew McConaughey, Maggie Q, Theo James, \\ Theo James, Dexter (TV series), Laker Girls, \\ Bayi Rockets, Arctic Monkeys, \\ Dulwich College Beijing\end{tabular} & \begin{tabular}[c]{@{}c@{}}Nicolas Cage, Alan Carr, \\ Liam Neeson, Dale Winton, \\ Craig Price (murderer), Manuel Neuer\end{tabular}                                                               \\ \hline
\end{tabular}
\end{adjustbox}
\end{table}

In Table \ref{tb:neighbors}, we further show article titles edited by three pairs of vandals. Each pair of vandals are close to each other in the embedding space.
The first row shows that two   vandals damage almost the same pages, which indicates vandals who edit the same pages are close to each other. The second row shows pages edited by the two vandals have common words in titles although the title names are different. This indicates our M-LSTM can discover the semantic collaborative behaviors on pages based on the user embeddings. The last row  shows that our M-LSTM can further identify vandals who damage the pages with similar subject areas although there are no any common words in the titles. This example shows the usefulness of incorporating page category information in our M-LSTM.
All above examples demonstrate that users who are close in the low-dimensional user embedding space have similar edit patterns. Therefore, analyzing user embeddings can help capture and understand collaborative vandal behaviors.

\subsection{Vandal Early Detection}

Our vandal early detection is achieved after each edit is submitted. Although our M-LSTM exploits revert status of the edit, we emphasize that the revert status is inspected by the ClueBot NG in a real time manner. Hence, our M-LSTM can be deployed for real time vandal detection.  We evaluate the vandal early detection on the 6427 users who have at least two edits in the testing dataset.
Table \ref{tb:early} shows the precision, recall and F1 of our M-LSTM on vandal early detection. We vary the threshold $\tau$ from 0.5 to 0.9. Similar to the results of vandal detection shown in Table \ref{tb:vandal}, with increasing $\tau$, a classifier with a higher threshold has more confidence about the prediction, resulting in a higher precision. On the contrary, the recall decreases with the increasing of $\tau$ because fewer users will be marked as vandals. The F1 score reaches the maximum with $\tau =0.9$. However, comparing with the results of vandal detection, the vandal early detection has a lower precision but much higher recall. This indicates that we lose some precision but achieve big recall when using partial edit information to do early vandal detection.

Table \ref{tb:early} further shows the average number of edits before the vandals were blocked by the administrators and the ratio of vandals who can be early detected over the whole testing dataset.
We can observe that the average number of edits keeps relatively steady while the threshold increases. Meanwhile, the ratio of early detected vandals has a significant decreasing while the threshold $\tau=0.9$. Note that the ratios of early detected vandals with thresholds from 0.5 to 0.8 are only a little lower than the recall values, which indicates that most of the vandals who are correctly detected are early detected. Overall, setting threshold $\tau=0.8$ will achieve a balance performance between vandal early detection and accurate prediction.

\begin{table}[]
\centering
\caption{Experimental results on precision, recall and F1 of vandal early detection, the average number of edits before the vandals are blocked, and the ratio of vandals who are early detected.}
\label{tb:early}
\begin{tabular}{|c|c|c|c|c|c|}
\hline
$\tau$ & Precision & Recall  & F1      & \# of Edits & \% of Early Detected \\ \hline
0.5    & 84.10\%   & 99.07\% & 90.97\% & 3.61        & 97.35\%              \\ \hline
0.6    & 84.96\%   & 98.99\% & 91.44\% & 3.60        & 96.87\%              \\ \hline
0.7    & 85.81\%   & 98.82\% & 91.86\% & 3.50        & 95.94\%              \\ \hline
0.8    & 86.88\%   & 98.76\% & 92.44\% & 3.59        & 93.34\%              \\ \hline
0.9    & 89.89\%   & 98.34\% & 93.93\% & 3.53        & 72.32\%              \\ \hline
\end{tabular}
\end{table}

\section{Conclusion and Future Work}
In this paper, we have developed a multi-source LSTM model to encode the user behaviors to user embeddings for Wikipedia vandal detection. The M-LSTM is able to simultaneously learn  different aspects of user edit information, thus user embeddings accurately capture the different aspects of user behaviors. Our M-LSTM achieves the state-of-the-art results on vandal detection. Furthermore, we showed that user embeddings are able to identify collaborative vandal groups with various patterns. Different from existing works which require a list of hand-designed features, our M-LSTM can automatically learn user embeddings from user edits.
The user embeddings  can be used for a variety of data mining tasks such as classification, clustering, outlier analysis, and visualization. Our empirical evaluation has demonstrated its potential for analyzing collaborative vandals and early vandal detection. In the future, we plan to incorporate into our M-LSTM more information about user edits, e.g., user-user relations and hyperlink relations among articles. These relations are modeled as graphs and can be naturally incorporated into M-LSTM by using network embedding approaches \cite{Perozzi2014Deepwalk,Tang2015Line}. We also plan to conduct comprehensive evaluations on collaborative vandal detection.

{\bf \noindent Repeatability.} Our software together with the datasets used in this paper are available at {\small \url{https://bitbucket.org/bookcold/vandal_detection}}.

\section*{Acknowledgments}
The authors would like to thank anonymous reviewers for their valuable comments and suggestions.
The authors acknowledge the support from the 973 Program of China (2014CB340404), the National Natural Science Foundation of China (71571136), the Basic Research Program of Shanghai (16JC1403000), and China Scholarship Council to Shuhan Yuan and Yang Xiang, and from National Science Foundation (1564250) to Panpan Zheng and Xintao Wu. This research was conducted while Shuhan Yuan visited University of Arkansas.



\end{document}